\documentstyle[prl,aps,psfig]{revtex}
\newcommand{\beq}{\begin{equation}}
\newcommand{\eeq}{\end{equation}}
\newcommand{\beqa}{\begin{eqnarray}}
\newcommand{\eeqa}{\end{eqnarray}}
\newcommand{\ax}{a^{\dagger}}
\newcommand{\cx}{c^{\dagger}}

\newcommand{\ssvec}{{\bf S}}

\newcommand{\sivec}{{\bf \sigma}}
\newcommand{\tbar}{\overline{t}}

\newcommand{\uni}{\rm{\bf{1}}}

\begin{document}
\draft

\twocolumn[\hsize\textwidth\columnwidth\hsize\csname
@twocolumnfalse\endcsname

\widetext 
\title{Interplay between Spin and Phonon Fluctuations in the 
double-exchange model for the manganites.}

\author{Massimo Capone}
\address{ Dipartimento di Fisica, Universit\`a di Roma ``La Sapienza'',
and  Istituto Nazionale per la Fisica della Materia (INFM) \\
Unit\`a  Roma 1, Piazzale Aldo Moro 2, I-00185 Roma, Italy}
\author{Sergio Ciuchi}  
\address{ Dipartimento di Fisica, Universit\`a de L'Aquila,
and  Istituto Nazionale per la Fisica della Materia (INFM) \\
Unit\`a de L'Aquila, Via Vetoio, I-67100 L'Aquila, Italy}
\date{\today}
\maketitle

\begin{abstract}

We present exact solutions, mainly analytical, for the two-site
double-exchange-Holstein model, that allow us to draw a complete picture of
the role of both phonon and spin quantum fluctuations in determining
the short-range correlations in the manganites.
We provide analytical solutions of the model for arbitrary electron-phonon
coupling and phonon frequency, for $S=1/2$ and for the classical spin limit
$S=\infty$, and compare these results with numerical diagonalization
of the realistic $S=3/2$ case.
The comparison reveals that the realistic case $S=3/2$ is not well described
by the classical spin limit, which is often used in literature.
On the other hand, the phonon fluctuations, parametrized by the
phonon frequency $\omega_0$, stabilize ferromagnetic phases
with respect to the adiabatic limit.
We also provide a complete analysis on the polaron crossover in this model.

\end{abstract}

\pacs{71.38.-k, 71.38.Ht, 75.50.Dd, 75.50.Ee }
]

\narrowtext

\section{Introduction}
\label{introduction}

It is  known from the 1950s that the double exchange mechanism 
\cite{zener,anderson,degennes} is at the basis of the magnetic properties
of the manganese perovskites R$_{1-x}$A$_x$MnO$_3$ (where R is a rare
earth element (e.g., La), and A is a divalent element like Sr or Ca).
In this compounds, the $d$-levels of each Mn ion host $4-x$ electrons. 
Three of them occupy the three low-lying $t_{2g}$ levels, 
with aligned spins due to the
Hund's rule. These electrons are basically localized, and only
the $1-x$ electrons in the $e_g$ level contribute to the
transport properties.
The discovery of the so-called ``colossal'' magnetoresistance \cite{cmr}
has originated an enormous revival of studies on these compounds,
both on the theoretical and the experimental side.
The systematic experimental investigations of the last few years have
underlined some weakness in the previous understanding,
unveiling a surprisingly rich phase diagram where a lot of
competing phases are stabilized by varying doping, temperature and
chemical nature of the dopants.
One of the most relevant new theoretical trends is
the suggestion that the transport properties cannot be fully
understood on the basis of the double exchange alone, and that
the interplay of this mechanism and a significant electron-phonon (e-ph)
interaction leading to a Jahn-Teller effect is the key for
the explanation of these properties \cite{dealone}.
The preminent role of e-ph effects has also been firmly established
experimentally by various groups and techniques \cite{expeph}.

The complex entanglement between charge, orbital and lattice
degrees of freedom represents a hard theoretical challenge, that
is far from being solved.
Many approximate solutions and numerical results have been proposed,
but the complexity of the phase diagram has naturally forced various
authors to many uncontrolled simplifications.
In this work we make a step back, and focus on a system
for which {\it analytical }exact results can be obtained.
Namely, we solve the double exchange model for a single electron
on two sites in the presence of a local Holstein e-ph coupling
\cite{holstein}.
Since we are interested in the relevant physics determining
the interplay between lattice and spin quantum fluctuations, we 
consider the simple Holstein model, instead of a more involved
Jahn-Teller coupling. This choice does not imply a loss of generality
since we are not discussing  the role of orbital degrees of freedom.

The two site system has been extensively studied as the minimal system able to
capture the key features of polaron formation from the point of view of ground
state \cite{Bellissard,Feinberg90,Alex,Firsov} as well as spectral
properties \cite{ranninger,ranninger-demello}.
Recently also the interplay between e-ph and e-e correlations have been studied
semi-analytically within the same model \cite{Acquarone}.
However, as far as magnetic properties are concerned, it is quite obvious
that a two-site system does not allow for long-range order and
phase transitions.
Nevertheless, it shows {\it short-range} (nearest neighbors)
correlations, that give substantial indications on the actual
long-range properties of the system, at least in strong
coupling. In the context of the
models for the manganites it is in fact believed that the
finite-size effects play a little role \cite{dagotto1}.
In the following we will define "first order transitions",
the level crossing between phases with different symmetry, and
"second order transitions" the continuous transitions.
We will discuss the relevance of our results to large systems
in the following.

Due to the simplicity of the model, we can give complete exact
phase diagrams without approximations.
One of our main results is a complete characterization of the
role of the quantum fluctuations of the core $t_{2g}$ spins.
In a microscopic model of the manganites, the core spins 
have $S=3/2$, and this value is usually thought to be large
enough to get rid of quantum fluctuations and treat them as classical
variables. We will explicitly test this assumption by comparing
the limiting cases $S=\infty$ (classical spins) and  the extreme
quantum case $S=1/2$, where the effect of quantum fluctuations is
maximum, with the realistic $S=3/2$ case.

Analogously, we will discuss the role of lattice quantum fluctuations,
releasing the adiabatic approximation on the phonon degrees of freedom.
The role of quantum phonon fluctuations is not trivial, as
already known for e-ph models alone \cite{csg,cdff,ccg}.
For $S=\infty$ and $S=1/2$ we give analytical exact solutions
of the model, exploiting an
exact analytical solution of the two-site Holstein model.
For $S=3/2$ we use standard exact diagonalization to solve the model.
Also in this case no approximation is introduced and all
the regimes are accessible.

A similar study has been reported in Ref. \cite{indiani},
where the two-site double exchange model for {\it classical spins}
is solved by perturbation theory around a variational reference
state obtained by a Lang Firsov canonical transformation.
Our work overcomes some limitations of Ref. \cite{indiani}, 
namely the classical spin limit.
Contrary to Ref. \cite{indiani}, we are also able to explore
the adiabatic regime $\omega_0 \ll t$, well beyond the region
in which the Lang Firsov result is a good reference state.

The paper is organized as follows: In Section \ref{methods} we introduce the
model and the methods for our analytical solutions; In 
Section \ref{results} we present the phase diagram of the model
and discuss the role of quantum fluctuations;
due to the complexity of the phase diagram
the discussion is divided in three subsections:
in the first the effect of the magnetic degrees
of freedom on the polaron crossover is considered;
in the second we discuss the effect of the e-ph interaction on 
the magnetic phase diagram of the model, and in the third subsection the
full phase diagram is presented.

In Section \ref{discussion} we discuss the relevance of our results
for larger size systems and for the experimental scenario.
Finally we give concluding remarks in Section \ref{conclusions}. 
\section{Methods of Solution}
\label{methods}
We consider the Holstein-Double Exchange model on a two-site cluster
for a single electron:

\beqa
\label{model}
H &=&-t \sum_\sigma(\cx_{1,\sigma} c_{2,\sigma} + \cx_{2,\sigma} 
c_{1,\sigma})+\nonumber\\ 
&&-J_H \sum_{i=1,2}
	\sivec_{i} \cdot \ssvec_i +J_1 \ssvec_1 \cdot \ssvec_2+\nonumber\\
&&-g (n_{1} - n_{2})(a+a^{\dagger})
+\omega_0 a^{\dagger}a,
\eeqa
where $\ssvec_i$ ($i=1,2$) is a local spin associated to the localized $t_{2g}$
electrons on each site, $c_{i,\sigma} (\cx_{i,\sigma})$ destroys
(creates) an electron of spin $\sigma$ on site $i$, $n_i = \sum_\sigma 
\cx_{i,\sigma}c_{i,\sigma}$ is the number operator on each site,
$\sivec_{i} = c^{\dagger}_{i\alpha}{\vec{\sigma}}_{\alpha\beta}
c_{i\beta}$ is the spin operator on each site (${\vec{\sigma}}$ are the Pauli matrices).
$a$ ($a^{\dagger}$) is the destruction (creation) operator for
a lattice distortion that couples to the difference of density
between the two sites. The lattice displacement $X$ is given 
by $X=\sqrt{\hbar/2m\omega_0}(a+a^{\dagger})$.
We could have started from a standard Holstein model with a phonon mode
per each site, coupled to the local density. 
It is in fact easy to show that, in this case, the symmetric combination 
 of the two phonon modes $A = 1/\sqrt{2} (a_1 + a_2)$  
couples to the total density, giving rise to a trivial term, and
the only term left is the one we introduced, where the 
phonon mode may be written in terms of the local ones as
$a = 1/\sqrt{2} (a_1 - a_2)$.

We explicitly consider, besides the 
hopping between the two sites and the Hund's rule term ($J_H$) that couples
ferromagnetically the conduction electrons to the localized ones,
an antiferromagnetic superexchange term ($J_1$) 
between the core electrons.
$J_H$ is always taken to be the largest energy scale, consistently
with the physics of the manganites.
This latter term, even though $J_1$ is significantly smaller than $J_H$,
has crucial importance on the
magnetic properties of the manganites\cite{manganoi}.
We also consider a Holstein coupling ($g$) between the electron
and a dispersionless mode of frequency $\omega_0$.

In this paper, we present the {\it exact} solution for the model (\ref{model}).
In particular for the classical spin case $S=\infty$ and the extreme
quantum case $S=1/2$ we provide {\it analytical} solutions for 
arbitrary values of both the electron-phonon coupling and of
the phonon frequency $\omega_0$, exploiting an exact
solution of the two-site Holstein model reported in Appendix \ref{cf}.
In the $S=3/2$ case, a numerically exact  solution by means of
exact diagonalization is instead presented.
In the following subsections, we describe the analytical solutions
for $S=1/2$ and $S=\infty$.

\subsection{Exact Solution for $S=\infty$}
Following Ref. \cite{anderson}, in the classical spins case we can
write 
\beq
\label{gs-Soo}
H(\theta)=J_1\cos{\theta} - H_H (\theta)+E_{Hund}
\eeq
where $E_{Hund}$ is the contribution of the Hund's term to the total energy and
$H_H$ is the hamiltonian of a 2-site Holstein model in which the 
hopping $t$ is replaced by $\overline{t}=t\cos(\theta/2)$. 
The canting angle $\theta$, that measures the relative
orientation of the core spins fully characterizes the magnetic
arrangement. If $\theta = 0$ the spins are aligned and a 
ferromagnetic ($FE$) state is found, whereas for $\theta = \pi$,
an antiferromagnetic state ($AF$) is found. Intermediate values
of $\theta$ describe canted ($CA$) states.
Therefore, the solution of the classical spins two-sites 
Holstein double-exchange model can be obtained by minimizing
on $\theta$, once the eigenvalues of $H_H$ are known, as shown in Appendix
\ref{cf}.
For $J_H \gg J_1$ the extremal condition then gives
\beq
\label{theta_c}
\sin (\theta/2)(-2 J_1\cos(\theta/2)-\frac{t}{2}\frac{\partial E_H}{\partial
\overline{t}})=0.
\eeq
which shows that the ferromagnetic state $\theta=0$ is always an extrema
of $E(\theta)$
\cite{note-extrema}. Then the transition from $FE$  to $CA$ state is 
of ``second order''. 
The critical coupling $J_1$ for this transition 
is given by the vanishing of the term in parentheses in Eq. (\ref{theta_c})
\beq
J_1^c = {-E_H^{kin} \over 4}
\label{SooCA}
\eeq
where $E_H^{kin} = t \partial E_H / \partial t$ 
is the kinetic energy of the Holstein model, i.e.,
the kinetic energy of the system with $J_1=0$ (Notice that $E_H^{kin}$
is a negative quantity). 
The effect of el-ph interaction on the 
$FE\rightarrow CA$ transition is therefore the substitution
$t\rightarrow-E_{kin}$.

Now let us consider the ``first order'' $FE\rightarrow AF$ transition. 
We have to compare the $FE$ and $AF$ energies obtained by Eq. (\ref{gs-Soo}) 
respectively with $\theta=0$ and $\theta=\pi$. 
The critical coupling is given by
\beq
J_1^c = {E_H(0)-E_H(t) \over 2}
\label{FE-AF_Soo}
\eeq
where $E_H(0)=-g^2/\omega_0$ is the energy of the atomic Holstein model.

\subsection{Exact Solution for $S=1/2$}
In this case, neglecting for the moment the phonon degrees of 
freedom, the electronic Hilbert space (including the core spins) 
 is in principle made by 16 states, that reduce to 8 if
the symmetry for inversion of all the spins is considered.
As shown in Appendix \ref{appendixb}, this problem can be
simplified, and the largest subspace to deal with is a 3$\times$ 3 sector,
but the remaining problem is still not trivial if we switch on 
the coupling with the phonons.
Fortunately, in the limit $J_H \gg t$, a further simplification occurs (also
shown in Appendix \ref{appendixb}), leading to the
possibility to express the eigenvalues of the model in terms
of the two-site Holstein model.
The details of the solution are reported in Appendix \ref{appendixb}.
In such a way, we can characterize the condition for the only possible
transition, i.e., the transition from $FE$ to $AF$ ground state.
The transition  is discontinuous and can 
be obtained from the comparison of the energies of the different phases,
that we compute in Appendix \ref{appendixb}.
The critical coupling for the $FE \to AF$ transition is then given by
\beq
\label{conds12}
J_1^c = {4(E_H(t/2)-E_H(t)) \over 3}
\eeq
where $E_H$ is the energy of the 2-sites Holstein model.
Contrary to the classical spin case
quantum spin fluctuations allow for non zero effective hopping $\tbar = t/2$
in the $AF$ phase.

\section{Results}
\label{results}
In this section, we present the phase diagrams of the model
(\ref{model}), with a particular emphasis on the interplay
between the role of phonon and spin quantum fluctuations (measured
respectively by $\omega_0/t$
and the value of the ``local'' spin of the $t_{2g}$ electrons $S$)
Due to the relatively large number of parameters that determine
the phase diagram, we organize the discussion of the results in subsections:
in the first subsection we discuss the polaron crossover in the different 
regimes, providing a unifying picture of the effect of both phonon and
spin fluctuations, and of electron-spin correlation, 
that generalizes in a consistent way 
the conditions for the small polaron crossover in the 
simplest  Holstein model\cite{csg,cdff,ccg}.
In the second subsection, we discuss the magnetic transitions 
occurring in our model. Both the nature of the magnetic phases and 
the relation between the magnetic state and the occurrence of 
polaronic behavior are strongly dependent on the value of the spin $S$.
Finally, in the last subsection, we present complete phase diagrams in
the $\lambda$-$J_1$ plane, where the role of lattice and magnetic 
degrees of freedom is highlighted.

\subsection{The polaron crossover}
\label{polaronsection}
Models with electron-phonon interaction
quite generally  exhibit a polaronic ground state when the coupling 
strength is large enough. The transformation of the free electron
into a small polaron is not a phase transition, but a continuous
crossover.
For the Holstein model and a single particle, it 
has been shown that the condition for the crossover significantly
depends upon the ratio between the phonon frequency $\omega_0$,
and the typical electronic energy scale $t$. In the adiabatic regime
$\omega_0 \ll t$, the crossover occurs for $\lambda =g^2/\omega_0t \simeq 1$,
whereas in the antiadiabatic regime $\omega_0 \gg t$, the crossover is 
controlled by the (purely phononic) variable $\alpha =g/\omega_0$
\cite{csg,cdff,ccg}.
As a matter of fact, the crossover coupling $\lambda_{pol}$
is pushed to larger values of $\lambda$
as the frequency is increased. Moreover, the crossover becomes smoother
and smoother as $\omega_0/t$ gets larger.

The conditions for a polaron crossover in the adiabatic and antiadiabatic
regimes can be understood on basic physical grounds.
In the adiabatic regime, the key condition is that a bound state can
be formed. The condition $\lambda > 1$ expresses this property, since
it just implies that the polaron binding energy $E_{pol}=g^2/\omega_0$
exceeds the kinetic energy of a free electron $\sim t$.
On the other hand, in the antiadiabatic limit, the electronic
energy scale $t$ is not the largest scale, and the polaron 
crossover is ruled by the condition $\alpha^2 > 1$, that corresponds
to the excitation of  a significant number of phonons 
(or, equivalently, to a sizeable lattice distortion).
The crossover conditions we have just described are based on simple, 
model independent, physical insights, and are therefore expected to 
basically hold, with some marginal changes, 
also for more complicated models like the double-exchange model we
are considering.

It must be noted that, since the formation of a polaron is not a phase 
transition with an associated broken symmetry, there is some
ambiguity in determining a physically sensible clear-cut criterion
for the polaron crossover.
In most previous studies, including that of Ref. \cite{indiani}, the crossover line has been drawn as the locus of
the points 
in which some relevant expectation values, like the electron-lattice
correlation function $1/N\sum_i\langle n_i X_i\rangle$
or the average number of phonons in the ground state change their behavior.
This kind of characterization has no problem in the adiabatic regime, where
the crossover is rather sharp, but it is more questionable in 
the antiadiabatic regime.
In this work we use a much more definite criterion, that is based
on a {\it qualitative} difference between polaronic and non-polaronic 
states. Namely, we study the (quantum) probability distribution function for
the displacement operator 
$P(X) = \langle 0\vert X\rangle \langle X\vert 0\rangle$, 
where $\vert 0 \rangle$ is the ground state wave function and $\vert X\rangle$
denotes the state with displacement $X$.
In the adiabatic limit $\omega_0 = 0$, the phonon degrees of freedom 
are described by classical variables, and no quantum fluctuations are
present. The solution of the model involves a minimization of the 
electronic groundstate as a function of $X$. 
As a result, the probability distribution is a single (or a few) 
$\delta$-function, centred at the values that minimize the energy.
More explicitly, if the system is not polaronic, a single value of 
$X$ minimizes the
energy, while in the polaronic
regime, two different minima are obtained.
The polaron crossover is then associated with the coupling value
in which a single $\delta$-function leaves place to 
two symmetric peaks.
As soon as the quantum fluctuations of the lattice are restored
by introducing a finite phonon frequency $\omega_0$, the
$\delta$-functions broaden, but the qualitative features do not
change. The polaronic regime is characterized by a {\it bimodal}
distribution,  
while the non-polaronic 
state present a {\it unimodal} distribution.
In the particular case of a single particle on two sites,
the polaronic regime presents two symmetric peaks at $X = \pm X_0$,
and the non-polaronic state in non-distorted, so that $P(X)$ is
peaked at $X=0$.

\begin{figure}
\centerline{\psfig{bbllx=80pt,bblly=200pt,bburx=510pt,bbury=575pt,%
figure=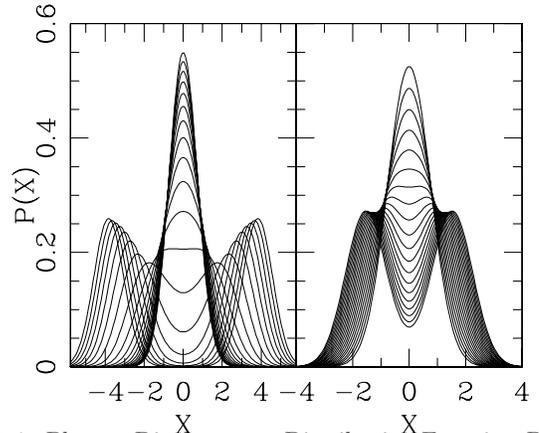,width=70mm,angle=0}}
\caption{Phonon Displacement Distribution Function $P(X)$ for the
two-site Holstein model for $\omega_0/t = 0.1$ and $\omega_0/t = 4$. 
The various lines in the two panels correspond to different values of
$\lambda$. For $\omega_0/t=0.1$ $\lambda$ ranges from $0$ to 1.4,
whereas for $\omega_0/t=4$ $0 < \lambda < 4$. 
\label{px_figure}
}
\end{figure}

Fig. \ref{px_figure} shows the evolution of $P(X)$ varying $\lambda$ 
in the two-site Holstein model for $\omega_0/t =0.1$ (representative
of the adiabatic regime) and $\omega_0/t=4$ (representative of the
antiadiabatic regime).
In both cases a smooth crossover occurs between a quasi-free electron 
state (unimodal distribution) and a polaronic state (bimodal distribution).
The figure also clearly shows that the crossover in indeed much
sharper in the adiabatic case than in the antiadiabatic one.
Furthermore, $P(X=0)$ in the polaronic region
rapidly vanishes soon after the crossover in 
the adiabatic case, while it stays finite in the antiadiabatic,
despite it is a local minimum in both cases.

It is worth to emphasize that that the monomodal to bimodal crossover of $P(X)$
has also been reported as signature to a crossover toward a 
"polaronic" state also in studies of
the Holstein model {\it in the thermodynamic limit} using Dynamical Mean Field
Theory \cite{FJS},\cite{Millis-Shraiman}.

\subsection{Magnetic Correlations}
\label{magneticsection}
In this section we discuss the behavior of the magnetic correlations
in the two-site double exchange model. 
It should be clear that a such a small system can not 
undergo phase transitions, and that we are only able to describe short
range correlations.
We parametrize the magnetic correlations between the two sites
by means of the scalar product $\langle {\bf S}_1 \cdot {\bf S}_2 \rangle$
between the core spins.

Since our interest in the model is motivated by the manganites, 
we will always assume that $J_H$ is the largest energy scale.
In the $S=\infty$ case we let $J_H$ go to infinity, and in the
finite spin case, we take $J_H = 10t$.

In the absence of electron-phonon coupling, and assuming that $J_H$ is the
largest energy scale, the direct antiferromagnetic exchange between the core
electrons determines the magnetic properties of the system.
For zero and small $J_1$, the spins are ferromagnetically
aligned due to the Hund's coupling.
Increasing $J_1$, antiferromagnetic correlations tend to appear.

\begin{figure}
\centerline{\psfig{bbllx=80pt,bblly=160pt,bburx=510pt,bbury=645pt,%
figure=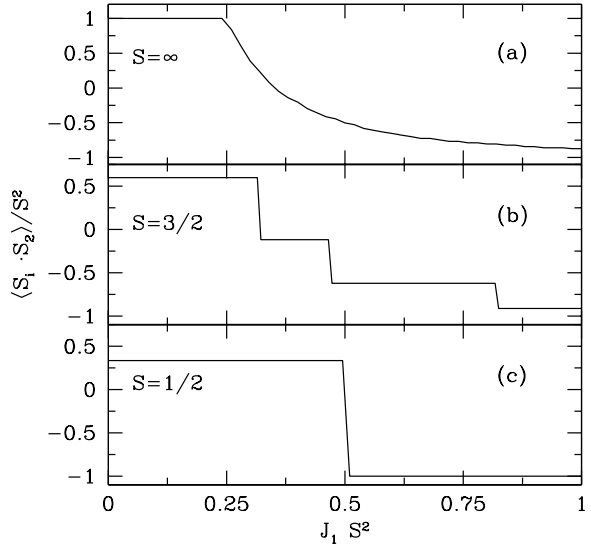,width=70mm,angle=0}}
\caption{$\langle {\bf S}_1 \cdot {\bf S}_1\rangle$ for $S=\infty, 3/2, 1/2$ (from top
to bottom) as a function of $J_1$.
\label{fig_magnetic}
}
\end{figure}

The nature of the spin correlations depends crucially on the value
of the spin $S$, since the latter rules the possible values of
$\langle {\bf S}_1 \cdot {\bf S}_2 \rangle$. 
More explicitly, in the classical spin case, 
$\langle {\bf S}_1 \cdot {\bf S}_2 \rangle = S^2cos(\theta)$,
where the canting angle $\theta$ between the spins is a continuous
variable. The ferromagnet continuously
evolves into a canted state as $J_1$ is enhanced. The canting angle 
asymptotically tends to $\pi$, corresponding to the antiferromagnetic
state,  as $J_1$ is enhanced.
Panel (a) in Fig. \ref{fig_magnetic} displays the
dependence of $\langle {\bf S}_1 \cdot {\bf S}_2 \rangle$ on $J_1S^2$ for
the classical spin case for $J_H = \infty$.

In the quantum case the total spin is given by   
\begin{equation}
\langle {\bf S}_1 \cdot {\bf S}_2 \rangle=
{1\over 2}(S(S+1) - S_1(S_1+1) - S_2(S_2+1))
\end{equation}
and assumes only a few values.
For $S_1=S_2=1/2$, $S=0$ and $1$ are the two only possible values.
For $S_1=S_2=3/2$, we can have four values ($S=0,1,2,3$).
It must be noted anyway, that the total spin operator $S^2$ does
not commute with the Hamiltonian (\ref{model}), so that the
energy eigenstates have no reason to be eigenstates of $S^2$.

An inspection to the results in the absence of 
electron-phonon coupling shows indeed that, for $S=1/2$,
the magnetic state abruptly varies, for $J_1 = J_1(FE-AF)$,
 from a {\it fully polarized}  ferromagnet ($FE$)
to an antiferromagnetic ($AF$) state, which is not fully polarized.
The transition is a level crossing between two states with 
different symmetry.
The exact value of $\langle {\bf S}_1 \cdot {\bf S}_2 \rangle$ in this 
state depends on both $J_H$ and $J_1$, even if the dependence on 
$J_1$ (for $J_1 > J_1(FE-AF)$) is really weak, as it appears in 
panel (c) of Fig. \ref{fig_magnetic}.

The $S=3/2$ case is strictly analogous to $S=1/2$, and shows
the fully polarized $FE$ state, followed by three different
combinations of the $AF$ states. Also in this case, the precise values
of $\langle {\bf S}_1 \cdot {\bf S}_2 \rangle$ in the three states
depend on $J_H$ and $J_1$, and the dependence on $J_1$ is really weak
in each region.
Moreover, the state with the largest negative correlation is really
close to the full antiferromagnet.
We label the two intermediate spin phases as canted 1 ($CA1$) and
canted 2 ($CA2$), and the ``most antiferromagnetic'' as antiferromagnetic
($AF$) {\it tout court}. The dependence of 
$\langle {\bf S}_1 \cdot {\bf S}_2 \rangle$ on $J_1$ is shown in 
Fig. \ref{fig_magnetic} (b).
We notice that the scale of $J_1$ associated with the change in the
magnetic structure is consistent with the experimental 
estimates \cite{expj1}.

\subsection{The phase diagram}
\label{phasediagram}
In this section we discuss the interplay between the 
magnetic properties and the e-ph coupling 
and finally determine the phase diagram of our model.
We tune the relevance of lattice and magnetic degrees of freedom, 
by varying the strength of the electron-phonon coupling $\lambda$, and of
the antiferromagnetic coupling between the core spins $J_1$.
Then we draw various phase diagrams in the $\lambda-J_1$ plane. Each of the
diagrams is characterized by the values of the spin $S$ and of the 
phonon frequency $\omega_0$, that parametrize the relevance of
quantum spin and lattice fluctuations respectively.
We consider the $S=\infty$, $S=3/2$ and $S=1/2$ cases, and 
$\omega_0/t = 0.1$ (adiabatic regime)  and $4$ (antiadiabatic regime).
Finally, we always assume that the Hund's rule coupling $J_H$ is the 
largest energy scale. In the classical case, we take $J_H = \infty$,
and in the quantum cases, we use $J_H = 10t$.

We denote ``first order transitions'' (level crossings between states with 
different symmetries) with full lines
and ``second order transitions'' (crossovers between phases with the
same symmetry) with dashed lines.

\subsubsection{The effect of $J_1$ on the polaron crossover}

In section \ref{polaronsection}, we have briefly described the
conditions ruling the polaron crossover in the Holstein model.
In the adiabatic regime $\omega_0/t \ll 1$, the condition
for a polaron ground state is that the 
polaron binding energy $E_{pol}=g^2/\omega_0$ is
larger than the free electron kinetic energy, measured by $t$.
It is quite natural to generalize this condition to the 
double-exchange model, at least when the polaron crossover occurs 
between two phases that share a common magnetic state.
In this case, we can replace the bare hopping $t$ by the 
``magnetically renormalized'' kinetic energy at $\lambda=0$
($\tbar$).
Thus the crossover condition is determined by the condition
\begin{equation}
\label{polcross}
\lambda_{mg} = {g^2 \over \omega_0 \tbar} \simeq 1.
\end{equation}

In the $S = \infty$ case the magnetic hopping is given by
\begin{equation}
\label{magnetichopping}
\tbar = t cos{\large (}{\theta\over 2}{\large )},
\end{equation} 
where $\theta$ is the canting angle between the $t_{2g}$ 
spins\cite{anderson,degennes}.
In the quantum cases $S=1/2$ and $S=3/2$, one can view the
canting angle as a quantized quantity, that can assume only a few
discrete values. We anticipate  that these are {\it not} the quantized
values of the semiclassical approximation.

Regardless the value of $S$ and $\omega_0/t$, 
for small values of $J_1$ the ground state 
is always ferromagnetic due to the Hund's rule. A crossover
occurs between a ferromagnetic itinerant electron and 
a ferromagnetic polaron. Within this region, 
the magnetic hopping is fixed to the free value $\tbar \equiv t$, 
and does not depend on $J_1$. As a result,
the model is completely equivalent to a two-site Holstein model,
and the crossover is associated with a vertical line in the $\lambda -J_1$
diagram, as shown in all the phase diagrams (Figs. \ref{ph1},\ref{ph2},
\ref{ph3},\ref{ph4},\ref{ph5},\ref{ph6}). 
The crossover value of $\lambda$ depends only on the
ratio $\omega_0/t$, and moves from the $\lambda \simeq 1$
in the extreme adiabatic limit, to $\lambda \simeq 1.2$  for
$\omega_0/t = 0.1$, to a substantially larger value ($\lambda \simeq 3.46$)
for $\omega_0/t = 4$, where the condition for the polaron
crossover is close to $\alpha^2 \simeq 1$ (that implies $\lambda \simeq 4$).

Increasing $J_1$, phases with antiferromagnetic correlation between the core
spins appear. The nature of this phases depends on the value
of $S$, as shown in section \ref{magneticsection}.
We start from the quantum cases, that present sharp level crossings 
at $\lambda=0$, where $\tbar$ sharply jumps following the magnetic
correlations shown in Fig. \ref{fig_magnetic}.
If we neglect the really weak dependence on $J_1$ of 
$\langle {\bf S}_1 \cdot {\bf S}_2 \rangle$ {\it within} a 
given magnetic phase,
the polaron crossover is controlled by the condition
(\ref{polcross}), where $\tbar$ is the value corresponding
to the actual magnetic phase.

The exact results obtained as described in section \ref{methods}
confirm this expectation,
and the polaron crossovers among phases with the same magnetic
correlation are in fact delimited by vertical dashed lines 
in all the diagrams for $S=1/2$ and $S=3/2$ (Figs. \ref{ph1},\ref{ph2},\ref{ph3},\ref{ph4}).
The value of the crossover coupling obviously changes in the different 
magnetic phases. The $FE$ state is the one with the 
largest kinetic energy due
to the double exchange mechanism, so that the critical $\lambda$
is the highest in this phase, and it decreases by decreasing the 
value of the magnetic correlation according to Eq. (\ref{polcross})
(see, e.g., Fig. \ref{ph1},\ref{ph2}).
Notice that the effect of the value of 
$\langle {\bf S}_1 \cdot {\bf S}_2\rangle$
on the crossover coupling is much more evident in the
adiabatic limit (Fig. \ref{ph1}), where the competition between the polaron 
energy and the kinetic energy rules the crossover, than in the antiadiabatic
regime (Fig. \ref{ph2}), 
where the electronic kinetic energy is not the most relevant 
quantity. In the extreme antiadiabatic limit $\omega_0/t \to \infty$
this dependence must completely disappear,
since the kinetic energy plays no role in the crossover.

The formation of polaron does not only occur as smooth crossover between
states with the same magnetic correlation. 
Indeed, in the $S=1/2$ case (see Fig. \ref{ph3}), 
if we continuously increase $J_1$, we have
that, between the vertical dashed lines corresponding to the polaronic 
crossovers within the $FE$ and the $AF$ phases, 
a first-order transition (level crossing)
occurs from a ferromagnetic non-polaronic state and an antiferromagnetic
polaron. The interplay between the localizing effect of both the 
e-ph and the antiferromagnetic magnetic interaction
strongly favors the $AF$ polaronic state with respect to the competing phases.

Similar level-crossings occur for all the finite-spin cases, with more
involved details depending upon the value of $S$ and $\omega_0$.

For example, in the antiadiabatic regime $\omega_0/t =4$, the
$FE-AF$ polaron transition occurs only in a narrow range of 
parameters (see Fig. \ref{ph4}, compared to the adiabatic case \ref{ph3}).
This is simply due to the fact that, in this regime, the crossover
values for $\lambda$ in the $FE$ and the $AF$ are very close.
In the extreme antiadiabatic limit $\omega_0/t \to \infty$ this region 
would indeed vanish.

In the richer $S=3/2$ case, the $CA1$ and $CA2$ states intrude between the
$FE$ and the $AF$ at weak e-ph coupling. 
In the adiabatic regime (see Fig. \ref{ph1}), no polaron
crossover occurs within the canted phases, and both these phases
undergo a first-order transition to the $AF$ polaron
Only the $FE$ and $AF$ phases display the usual polaron crossover.
In the antiadiabatic case, besides the aforementioned reduction of
the regions in which the polaron formation becomes first-order,
canted polaronic states are stabilized by the phonon quantum fluctuations.
(The ``critical'' frequency above which canted polaronic states appear
is $\omega_0/t \simeq 1$).

In the $S=\infty$ case, where canted phases with a continuous canting angle
are stable, the polaron crossover is not represented by a vertical line,
since the kinetic energy is a continuous function of $J_1$
with $\tbar = t cos \bar{\theta(J_1)}$, and $\bar {\theta(J_1)}$
is the value of the canting angle at $\lambda =0$. 
Again, all results are consistent with the condition
(\ref{polcross}).
In this case, the phonon fluctuations play a somewhat qualitative
role. In the extreme adiabatic limit the $CA$ state undergoes
a level crossing to the $AF$ polaronic state. The e-ph interaction
and the antiferromagnetic coupling $J_1$ cooperate to
stabilize the $AF$ polaron without forming a canted polaron.
As soon as we introduce a finite, but small $\omega_0/t$, a tiny slice
of a canted polaronic phase appears to bridge between the
canted state and the antiferromagnetic polaron.
In the antiadiabatic regime, the huge quantum fluctuations strongly favor
a canted polaronic state, and the antiferromagnetic polaron appears
only for $\lambda > 10$ and $J_1 > 2$.

The above results show that the effect of the magnetic correlations
on the small polaron crossover is influenced by the spin quantum
fluctuations. 
In particular, the $S=3/2$ case, which is relevant to the manganites is
not qualitatively similar to the classical spin case, that is usually
considered for simplicity.
Many features of the $S=3/2$ case are in fact direct consequences
of the quantum nature of the spins, and are similar to the simplest
quantum case $S=1/2$.

\subsubsection{The effect of e-ph coupling on magnetic transitions}

In this section we analyze how the various magnetic transitions
described in Section \ref{magneticsection} are influenced by the
e-ph interaction.
In the previous section we have found analytical results for the
transition from $FE$ to $AF$ states  for $J_H \to \infty$.
The expressions, given by Eqs. (\ref{FE-AF_Soo}) 
and (\ref{conds12}), can be recast in the common form
\begin{equation}
\label{criterione}
J_1^c S^2 = {E_H(AF) - E_H(FE) \over {\langle {\bf S}_1 \cdot {\bf S}_2\rangle_{FE} \over S^2} - {\langle {\bf S}_1 \cdot {\bf S}_2\rangle_{AF} \over S^2}},
\end{equation} 
where $E_H(AF)$ ($E_H(FE)$) is the energy of the Holstein model for
the antiferromagnetic (ferromagnetic) spin alignment, and
$\langle {\bf S}_1 \cdot {\bf S}_2\rangle$ in the different magnetic phases
is computed at $g=0$.
In particular, the evaluation of $E_H$ for a given phase simply amounts
to find the groundstate of the Holstein model where the bare 
hopping $t$ is replaced by $\tbar$.
This relation also holds for the case of $S=3/2$, once the proper
value for the kinetic energy in the antiferromagnetic phase
$\tbar = t/4$ is used.

Eq. (\ref{criterione}) results from the competition between 
the magnetic energy balance, controlled by $J_1$, and 
the ``polaronic'' energy, i.e., the energy resulting from the e-ph
coupling. 

Since the $AF$ phase has always a smaller hopping with respect to
the $FE$ phase, 
the first qualitative effect of the e-ph interaction is to lower the 
energy of the antiferromagnetic phase with respect to the
ferromagnetic one, therefore favoring antiferromagnetism.
The region of stability of the $FE$ phase is always shrunk   
by increasing $\lambda$.
More generally, the e-ph interaction favors phases with smaller 
values of $\langle
\bf{S_1} \cdot \bf{S_2}\rangle$, so that the boundaries of the 
different magnetic phases are always marked by downward curves
in the $J_1$-$\lambda$ plane.

In the limit of small $t$, the energy difference in the numerator
of (\ref{criterione}) reduces to the pure kinetic energy of the
Holstein model close to the atomic limit.
Defining $\gamma = \tbar/t$, we can write
\beqa
&\lim_{t\to 0} &(E_H(\gamma t) - E_H(t)) = \nonumber\\ 
& = &\lim_{t\to 0} (1-\gamma) t
{E_H(t+(1-\gamma) t) - E_H(t) \over {(1-\gamma) t}}= 
\nonumber\\
& = & (1-\gamma) t{\partial E_H(t) \over \partial t} = -E_H^{kin}(1-\gamma).
\eeqa

In the small $t$ limit the condition (\ref{criterione}) reduces
then to 
\begin{equation}
J_1^c = \kappa(S)(-E_H^{kin}),
\end{equation}
where $\kappa(S)$ is a constant that contains the factor $1-\gamma$, and
depends on the value of
the spin $S$ and on the transition under consideration, such that,
in the absence of e-ph interaction the condition is simply
$J_1^c = \kappa(S)t$.
This latter relation is analogous to Eq. (\ref{polcross}),
since the effect of the e-ph interaction results in the substitution
$t \to -E_H^{kin}$, but, contrary to that, it is valid only for $t\to 0$.

This result gives valuable informations about the role of
the retardation effects in the e-ph coupling.
In general terms, the interaction mediated by the phonons is in fact
retarded, and becomes instantaneous only if 
$\omega_0/t \to \infty$. 
In such a limit, as we have shown, the kinetic energy rules
the magnetic transitions. As soon as the approximation
$t \to 0$ is released, the retardation effects imply that
Eq. (\ref{criterione}) must be used. Notice that 
$E_H(AF) - E_H(FE) = -E_H^{kin} +\Delta$, where $\Delta$ is 
the difference between the e-ph interaction energies in th
two phases, and turns out to be always positive.
The overall effect of the retarded e-ph interaction is therefore
to reduce the stability of $FE$ phases with respect to $AF$ and $CA$ phases.

In the $S=3/2$ case all the magnetic transitions are of ``first
order'', as described in section \ref{magneticsection}, so that
similar arguments can be applied, and Eq. (\ref{criterione}) 
allows to compute the transition coupling, once the energy of the Holstein 
model and the magnetic energy of the appropriate phases are known.

The situation is different only for the ``second order'' transition
between the $FE$ and $CA$ phases in the classical spin case.
In this case the continuity of the transition implies that the
energy difference between the phases is infinitesimal, 
so that $E_H^{kin} = -t\partial E_H(t)/\partial t$ rules the 
transition not only for small $t$, but for arbitrary values
of $t$, as shown by Eq. (\ref{SooCA}).
This preliminary analysis suggests a really important difference
between the classical spin limit $S=\infty$ and the 
quantum $S=3/2$ case. 

Now we can give some description and interpretation of the exact 
phase diagrams in light of the above analysis.

\begin{figure}
\centerline{\psfig{bbllx=80pt,bblly=200pt,bburx=510pt,bbury=575pt,%
figure=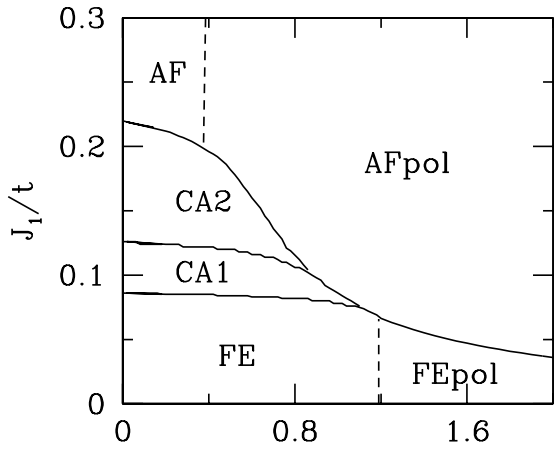,width=70mm,angle=0}}
\caption{Phase diagram for $S=3/2$ and $\omega_0/t=0.1$.
\label{ph1}
}
\end{figure}

\begin{figure}
\centerline{\psfig{bbllx=80pt,bblly=200pt,bburx=510pt,bbury=575pt,%
figure=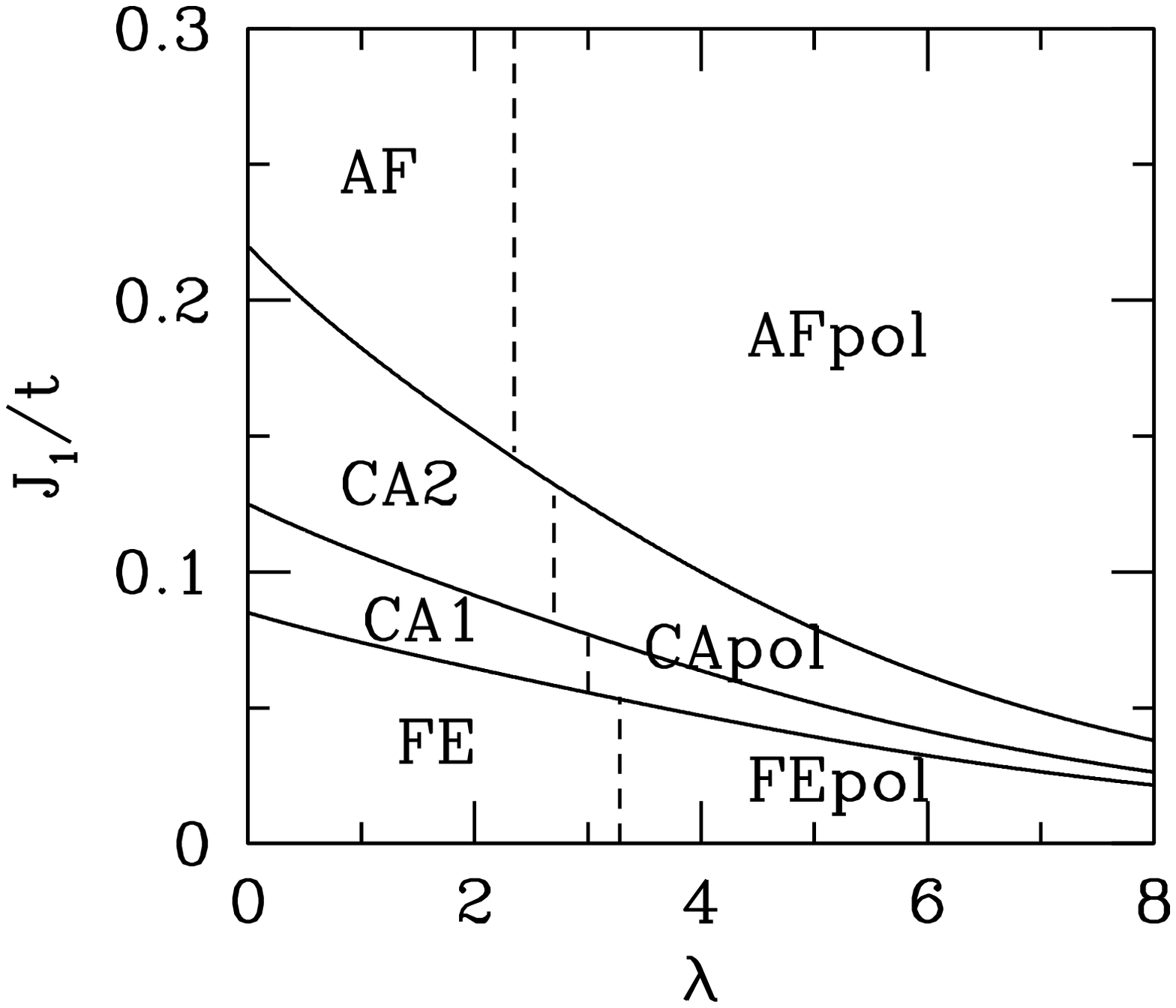,width=70mm,angle=0}}
\caption{Phase diagram for $S=3/2$ and $\omega_0/t=4$.
\label{ph2}
}
\end{figure}

\begin{figure}
\centerline{\psfig{bbllx=80pt,bblly=200pt,bburx=510pt,bbury=575pt,%
figure=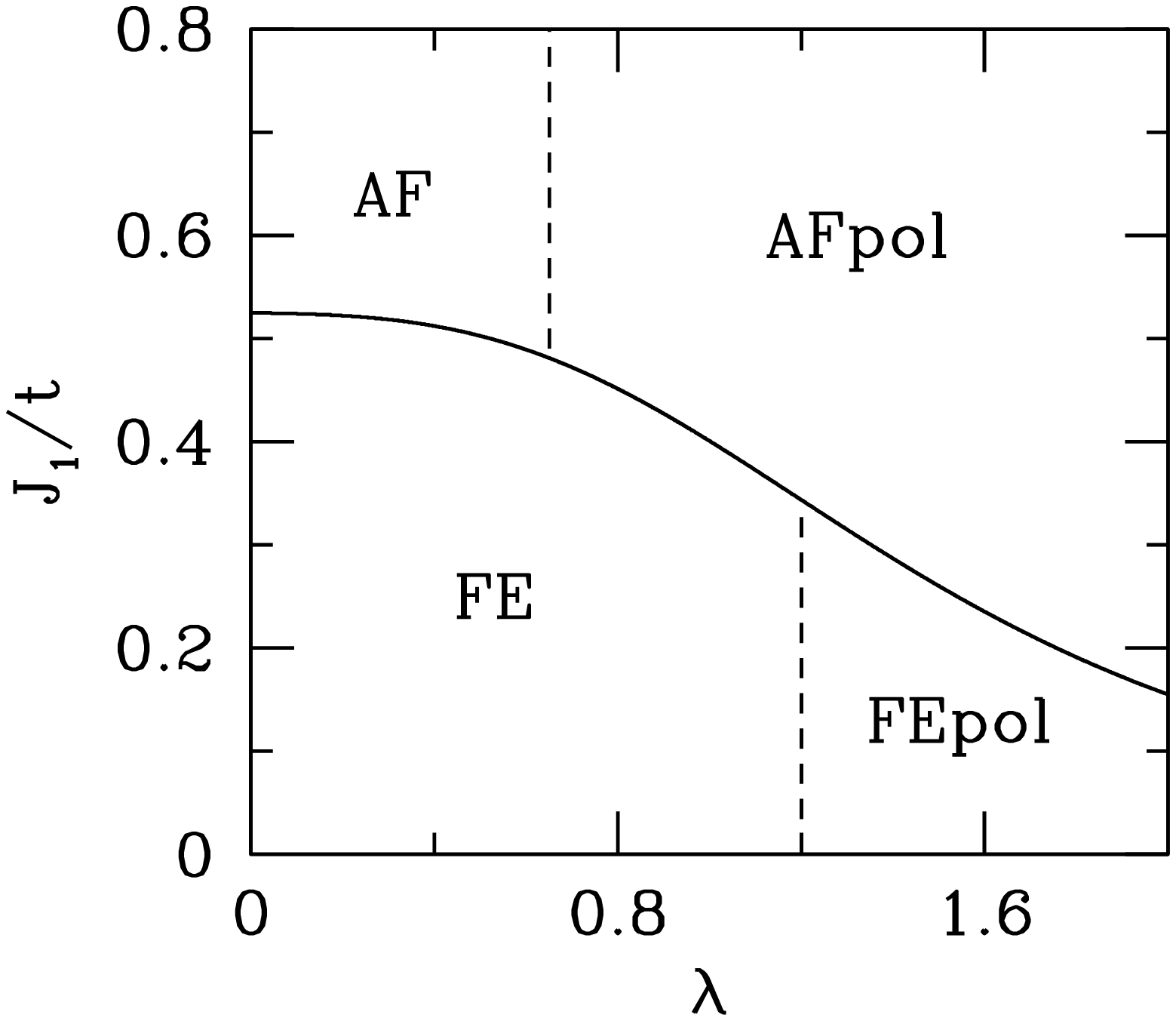,width=70mm,angle=0}}
\caption{Phase diagram for $S=1/2$ and $\omega_0/t=0.1$.
\label{ph3}
}
\end{figure}

\begin{figure}
\centerline{\psfig{bbllx=80pt,bblly=200pt,bburx=510pt,bbury=575pt,%
figure=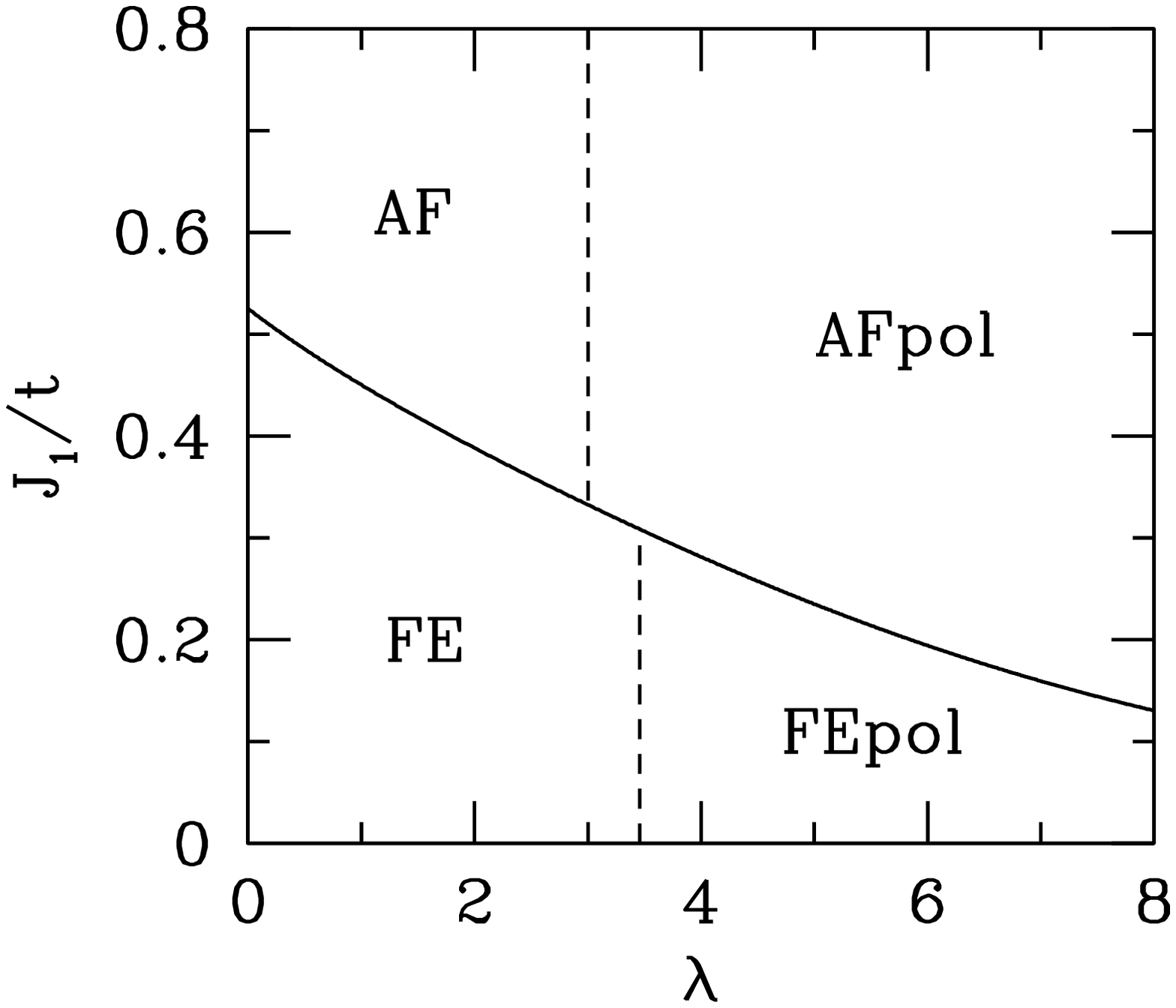,width=70mm,angle=0}}
\caption{Phase diagram for $S=1/2$ and $\omega_0/t=4$.
\label{ph4}
}
\end{figure}

\begin{figure}
\centerline{\psfig{bbllx=80pt,bblly=200pt,bburx=510pt,bbury=575pt,%
figure=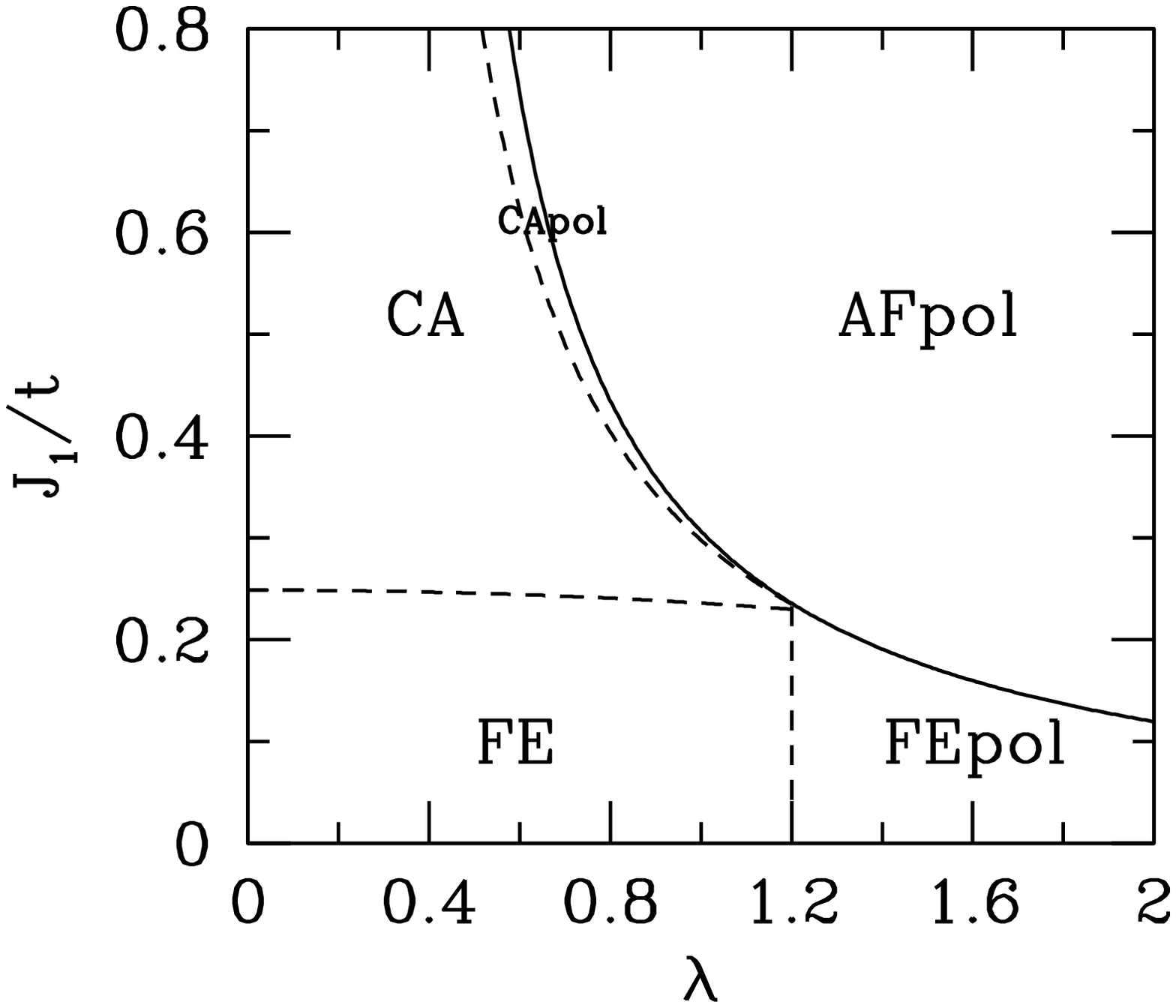,width=70mm,angle=0}}
\caption{Phase diagram for $S=\infty$ and $\omega_0/t=0.1$.
\label{ph5}
}
\end{figure}

\begin{figure}
\centerline{\psfig{bbllx=80pt,bblly=200pt,bburx=510pt,bbury=575pt,%
figure=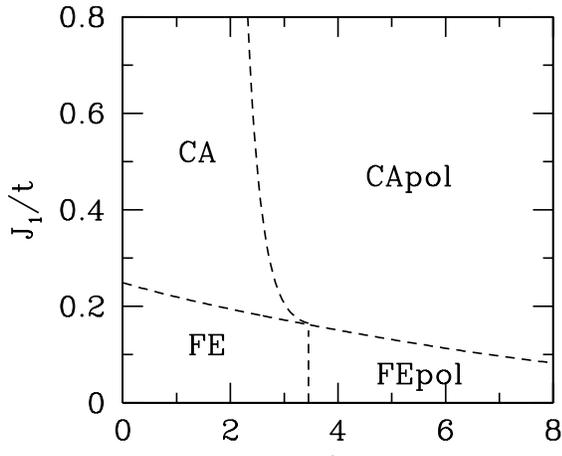,width=70mm,angle=0}}
\caption{Phase diagram for $S=\infty$ and $\omega_0/t=4$.
\label{ph6}
}
\end{figure}

In the adiabatic limit $\omega_0/t =0$, for finite value of $S$, 
and for $\lambda < \lambda_{pol}$, the energy of the e-ph model alone 
does not depend upon $\lambda$, so that the relative
stability of the various magnetic phases is in turn expected to be
$\lambda$-independent.
The magnetic transitions are therefore associated with horizontal lines in the
$\lambda$-$J_1$ plane.
If we introduce phonon fluctuations, the kinetic energy depends 
upon $\lambda$ also
before the crossover, and the boundary between the magnetic states 
acquires a finite slope, that becomes larger and larger by increasing
$\omega_0/t$.
This behavior can be easily seen comparing  Fig. \ref{ph1} with Fig. \ref{ph2} 
(or \ref{ph3} with \ref{ph4}).
The case of the classical spin variables, where the
magnetization is a continuous variable, can be understood in similar terms.
The crossover between the $FE$ and the $CA$ phase in fact closely follows the
curve obtained using Eq. (\ref{SooCA}).

As discussed previously,  the phonon fluctuations
always favor $FE$ phases with respect to 
$AF$ and $CA$ phases, as it can be seen comparing the phase diagrams
for $\omega_0/t=0.1$ with the corresponding with $\omega_0/t=4$
(at the same value of $\lambda$). 
In we increase the quantum fluctuations of the phonons, by 
increasing the phonon frequency $\omega_0$, the retardation
effects are decreased, so that the localization of the electrons
is made more difficult. An enhanced mobility of the
electron results in an enhanced stability of the $FE$ phases
due to the double exchange mechanism.


\section{Relevance for larger systems}
\label{discussion}
In this section we discuss the relationship between the two-site
model and larger systems.
There are not indeed so many solutions of the double exchange model in the 
presence of e-ph coupling, and most of them are forced to consider 
the classical limit for both the phonons and the core spins.
In the case of numerical calculation the main problem in dealing
with quantum phonons and spins is the enlargement of the Hilbert space
(that becomes infinite for the case of phonons).
This problems are even harder if the orbital degrees of freedom are taken
into account.
In Ref. \cite{dagotto1}, a numerical analysis of the double-exchange model
in presence of a Jahn-Teller coupling is performed for the
three-dimensional structure of the actual compounds, and for a number
of electrons corresponding to the stoichiometric LaMnO$_3$.
The three-dimensional structure and the realistic hopping matrix
elements allow for different antiferromagnetic arrangements of the spins
besides the ferromagnetic state. In particular, realistic values
of the parameters give rise to the A-type antiferromagnetism found
in the experiments \cite{dagotto1,manganoi}.
Of course our two-site system is not enough to distinguish among
the various kind of antiferromagnetic orderings, but the phase
diagram in Fig. 2(b) of Ref. \cite{dagotto1} and our
phase diagram for classical spins and for small phonon frequencies
(Fig. \ref{ph5}) are rather similar.
The shape of the curves separating the $FE$ phase to the ``non-$FE$'' phases
look very similar, and the crossover line between a $FE$ with no distortions
and a $FE$ polaronic phase is very close to the one between the $FE$ and
the $FE(JT)$ phase. The agreement is not only qualitative, but also 
quantitative as it can be checked by direct comparison.

The capability of the two-site cluster to capture the physics
of both the polaron crossover and the magnetic transitions in a more
realistic system is consistent with previous claims (see, e.g. \cite{dagotto1})
of the irrelevance of finite-size effects for these systems.

The $x=0.5$ system has been considered in 
Refs. \cite{dagotto2,denis}.
Also in this case, the phase diagram is quite similar to Fig. \ref{ph5}:
A first-order transition separates a $FE$ state to non-$FE$ states,
and within the $FE$ phase a vertical line denotes a crossover associated 
to small polaron formation (or Jahn-Teller effect).

The two-site cluster gives therefore results really close to 
larger systems when the latter are available.
This suggests that our results are generically representative
of larger sizes, also for parameters that are presently unaccessible
to numerical simulations.
We emphasize that this agreement is not accidental and can be easily 
rationalized.
As mentioned in the introduction, the small polaron crossover is 
well represented by a two-site cluster. This is essentially due
to the extreme short-range character of the polaronic state.
The capability of the two-site cluster to describe the crossover 
from a delocalized state and a polaronic state
is well documented.
It must be noticed, however, that our small cluster can not 
determine whether the polaronic and the delocalized states are
Fermi liquids or not.

As far as the magnetic properties are concerned, the two-site
cluster has no long-range correlation, but only short-range
correlations.
Nevertheless the features related to local 
interactions such as the ones in Hamiltonian (\ref{model}) are
well represented. In this spirit we expect the results of
our calculations to be somehow related to a Dynamical Mean
Field Theory, where the local quantum fluctuations are
exactly taken into account, while the spatial correlations
are frozen \cite{revdmft}.
Similarly to the Dynamical Mean Field Theory, our exact solutions
faithfully reproduce the physics of the model if the local and 
short-range effects are dominant, as it is the case for the
manganites.
More specifically the one electron solution of the two site Holstein model
(see Appendix \ref{cf}) is surprisingly similar to the
one electron solution of the Holstein model for an infinite lattice
within Dynamical Mean Field Theory. Looking at a given site, the role played by
the "effective quantum medium" of Dynamical Mean Field Theory is here simply
played by the other site.

\section{Conclusions}
\label{conclusions}
In this work the two-site double exchange model for an electron coupled with
phonons is solved exactly for an extremely wide range of
parameters and physical regimes.
For $S=1/2$ and $S=\infty$ we give an analytical exact solution
for arbitrary e-ph coupling and phonon frequency.
For $S=3/2$ the solution is obtained through standard numerical techniques.
The availability of these solutions allows us to study the 
effect of both phonon and spin quantum fluctuations,
and of their mutual interplay.

This study, though limited to the extreme small size of the two-site
cluster, is shown to be a good description of larger systems, 
since the relevant physics involves local and short-ranged quantities.

One of our main results is a complete characterization of the
effect of the double exchange and of an antiferromagnetic coupling
between the $t_{2g}$ spins on the small polaron crossover.
In this regard, we give an
analytical estimate for the crossover coupling, given by $\lambda_{mg}=
g^2/\omega_0 \tbar \simeq 1$, where $\tbar$ is the kinetic
energy renormalized by the magnetic effects.

From a complementary point of view, we considered in detail the effect
of the e-ph interaction on the magnetic properties of the system.
In this case, we give an analytical condition for the 
magnetic transitions in the presence of a finite $\lambda$,
given by Eq. (\ref{criterione}).
This relation can be simplified in the limit $t\to 0$, where 
there is no retardation effect.
The comparison between the general case and the atomic limit allows
us to quantitatively describe the role of retardation in stabilizing $AF$
(or $CA$) phases.

A comparison of the realistic $S=3/2$ case with the classical spin case
$S=\infty$ shows that this latter approximation does not reproduce
some qualitative features of the phase diagram.
A proper study of the manganites should therefore take into account
the quantum nature of the core spins.

\begin{acknowledgments}

We acknowledge useful discussions with D. Feinberg and C. Castellano.

\end{acknowledgments}

\appendix

\section{The exact solution of the two-site Holstein model}
\label{cf}
In this appendix we sketch the solution trough a continued fraction
expansion of the two site Holstein model. 
A continued fraction solution has been already reported in literature
\cite{ottici} for a related model in the field of quantum
optics. Here we derive the continued fraction expansion for eigenvalues and
eigenvectors of the two site Holstein model using a different method.

The Holstein model for an electron on two sites can be written
in a pseudo-spin representation in terms of the Pauli matrices 
\beq
H_H = \omega_0 \uni \ax a-g\sigma_z (a+\ax)-t\sigma_x
\eeq
where $\uni$ is the unity matrix.
The hamiltonian can be diagonalized in the electron subspace using a
transformation introduced in Ref. \cite{Bellissard}
\beq
U=\frac{1}{\sqrt{2}}\left (\begin {array}{cc}
1&(-)^{\ax a}\\\noalign{\medskip}
-1&(-)^{\ax a}
\end {array}\right )
\label{U-2sH}
\eeq
and the property
\beq
(-)^{\ax a}(a+\ax)(-)^{\ax a}=-(a+\ax)
\label{UX}
\eeq
we obtain for $\tilde{H}_H=UH_HU^{-1}$
\beq
\tilde{H}_H = \left (\begin {array}{cc} 
\tilde{H}_H(t)&0\\\noalign{\medskip}
0&\tilde{H}_H(-t)
\end {array}\right )
\eeq
where
\beq
\tilde{H}_H(\pm t)=\omega_0 \ax a-g (a+\ax) \mp t (-)^{\ax a}.
\label{toy-model}
\eeq
In each block we have a purely phononic hamiltonian $\tilde{H}_H(\pm t)$.
The eigenvalues and eigenvectors can be determined 
by continued fraction solution for the resolvent between
$|m\rangle$ and $|n\rangle$-phonon states
\beq
G^{\pm}_{m,n}(\omega)=\langle m|\frac{1}{\omega-\tilde{H}_H(\pm t)}|n\rangle
\eeq

Using 
\beq
\frac{1}{\omega-H}=\frac{1}{\omega-H_0}+\frac{1}{\omega-H_0}H_I\frac{1}
{\omega-H}
\eeq
with $H_0=\omega_0 \ax a \mp t (-)^{\ax a}$ and $H_I=-g (a+\ax)$ we get the
recursion
\beqa
G^{\pm}_{m,n}(\omega)& = & \delta_{m,n}G^{\pm}_0(\omega-n\omega_0) \nonumber\\
&& -g \sum_p G^{\pm}_0(\omega-n\omega_0) X_{n,p}G^{\pm}_{p,n}(\omega)
\eeqa
where $X_{n,p}=\langle n|a+\ax|p\rangle$. This tri-diagonal recursion can
be solved for the diagonal elements through a continued fraction solution
\cite{book}
\begin{equation}
\label{Gnn}
G^{\pm}_{n,n}(\omega)={1 \over\displaystyle \omega-n\omega_0 \pm t
-\Sigma_{em}-\Sigma_{abs}}
\end{equation}
where
\begin{equation}
\label{Sigma_abs}
\Sigma_{abs}=
{\strut ng^2 \over\displaystyle \omega+\omega_0 \mp t-
{\strut (n-1)g^2 \over\displaystyle \omega+2\omega_0 \pm t-
{\strut (n-2)g^2 \over\displaystyle \ddots -
{\strut g^2 \over\displaystyle \omega+n\omega_0 +(\mp)^n t}}}}
\end{equation}
and
\begin{equation}
\label{Sigma_em}
\Sigma_{em}=
{\strut (n+1)g^2 \over\displaystyle \omega-\omega_0 \mp t-
{\strut (n+2)g^2 \over\displaystyle \omega-2\omega_0 \pm t-
{\strut (n+3)g^2 \over\displaystyle \omega-n\omega_0 \mp t- 
\cdots}}}
\end{equation}
At zero temperature the Green function of the two site Holstein model
defined as
\beq
G_{i,j}(\omega)=-i\langle0|T c_i(t) c^\dagger_j(0)|0\rangle
\eeq
can be expressed as
\beqa
G_{1,1}(\omega)=\frac{1}{2}(G^{+}_{0,0}(\omega)+G^{-}_{0,0}(\omega))\\\nonumber
G_{1,2}(\omega)=\frac{1}{2}(G^{+}_{0,0}(\omega)-G^{-}_{0,0}(\omega))
\eeqa
therefore $G^{\pm}_{0,0}$ are the $k$-space propagators whose poles determines
the bonding and anti-bonding eigenvalues of $H_H$. The residues of the lowest
energy pole of $G^{\pm}_{n,n}$ determine the square of the  projection of the
phonon ground state on the $|m\rangle$-state $b^{\pm}_m$. Let us write Eq.
(\ref{Sigma_em}) for $n=0$ in a recursive fashion
\beq
\Sigma^{\pm}_p = \frac{pg^2}{\omega-p\omega_0+(\mp)^p t-\Sigma^{\pm}_{p+1}} 
\eeq
where $\Sigma^{\pm}_1$ is the continued fraction of Eq. (\ref{Sigma_em})
therefore the equation which gives the eigenvalue of the two site 
Holstein model is
\beq
\label{pole}
\omega\pm t-\Sigma^{\pm}_1=0.
\eeq

By linearizing this recursion around the $m$-th solution $E_m$ 
of Eq. (\ref{pole}) letting 
$z^{\pm}_p=\partial{\Sigma^{\pm}_p}/\partial{E_m}$ we get
\beq
z^{\pm}_p=(z^{\pm}_{p+1}-1)\frac{pg^2}{E_m-p\omega_0+(\mp)^p t -\Sigma^{\pm}_{p+1}}
\eeq.
Finally the coefficient $b^{\pm}_m$ is given by
\beq
b^{\pm}_m=\sqrt{\frac{1}{1-z^{\pm}_1}}
\eeq.
 
The (quantum) probability distribution function for
the displacement operator can be determined using the harmonic oscillator wave
functions $\Psi_n(X)$ as
\beq
P(X) = \sum_{m,n}\left [ (b^{+}_m)^*b^{+}_n+(b^{-}_m)^*b^{-}_n
\right ] \Psi^*_m(X)\Psi_n(X) 
\eeq

\section{The Exact Solution of the Two-site Holstein Double-Exchange model
for $S=1/2$}
\label{appendixb}
Let us start form  the case $g=0$.
We choose the following basis set labelling the states according to
the total spin $S_{tot}= S + s$ where $S$
is the spin of the Mg$^{3+}$ ion and $s$ that of the $e_g$ electron:
We have two states in the $S=3/2$ sector:
\begin{itemize}
\item{$|A\rangle=|\Uparrow\Uparrow\rangle|\uparrow .\rangle$,$|A'\rangle=|\Uparrow\Uparrow\rangle|. \uparrow\rangle$}
\end{itemize}
and six states in the $S=1/2$ sector:
\begin{itemize}
\item{$|B\rangle=|\Uparrow\Downarrow\rangle|\uparrow .\rangle$,$|B'\rangle=|\Downarrow\Uparrow\rangle|. \uparrow\rangle$}
\item{$|C\rangle=|\Downarrow\Uparrow\rangle|\uparrow .\rangle$,$|C'\rangle=|\Uparrow\Downarrow\rangle|. \uparrow\rangle$}
\item{$|D\rangle=|\Uparrow\Uparrow\rangle|\downarrow .\rangle$,$|D'\rangle=|\Uparrow\Uparrow\rangle|.\downarrow \rangle$},
\end{itemize}

where $\vert\Uparrow\rangle$ ($\vert\Downarrow\rangle$) represent
an up (down) spin state for the core spins and $\vert\uparrow\rangle$ 
($\vert\downarrow\rangle$) are the same for the $e_g$ electrons.
The Hamiltonian is invariant for flipping of all the spins so these are all
the states we need.
The states $|A\rangle$ and $|D\rangle$ have a $FE$ character while the states $|B\rangle$ and $|C\rangle$ have
$AF$ character. The $S=3/2$ subspace, spanned by the
combinations of $|A\rangle$ and $|A'\rangle$, decouples from the
other states even in the presence of e-ph phonon interaction

If we consider the symmetric and antisymmetric combinations
\begin{eqnarray}
|A^{\pm}\rangle=\frac{1}{\sqrt{2}}(|A\rangle\pm|A'\rangle)\nonumber\\
|B^{\pm}\rangle=\frac{1}{\sqrt{2}}(|B\rangle\pm|B'\rangle)\nonumber\\
|C^{\pm}\rangle=\frac{1}{\sqrt{2}}(|C\rangle\pm|C'\rangle)\nonumber\\
|D^{\pm}\rangle=\frac{1}{\sqrt{2}}(|D\rangle\pm|D'\rangle),
\end{eqnarray} 

in the absence of e-ph interaction the symmetric and antisymmetric
sectors are decoupled so that the part of the 
Hamiltonian matrix which pertains to Hund and antiferromagnetic 
interactions consists of 3 blocks:

\beq
H_{3/2} = 	\left ( \begin{array}{cc}
			\frac{J_1-J_H}{4}-t&0\\
			0&\frac{J_1-J_H}{4}+t
			\end{array} 
		\right)
\label{H_A}	
\eeq

\beq
H_{1/2,+} = 	\left ( \begin{array}{ccc}
			-\frac{J_1+J_H}{4}&\frac{J_1}{2}-t&0\\
			\frac{J_1}{2}-t&-\frac{J_1-J_H}{4}&-\frac{J_H}{2}\\
			0&-\frac{J_H}{2}&\frac{J_1+J_H}{4}-t\\
			\end{array} 
		\right)
\label{H_piu}
\eeq

the last block $H_{1/2,-}$ can be obtained from Eq. (\ref{H_piu}) by the
substitution $t\rightarrow -t$.
The e-ph   
 matrix elements couple the subspaces
($A^+$,$B^+$,$C^+$,$D^+$) and ($A^-$,$B^-$,$C^-$,$D^-$).
The subspace spanned by $|A^{\pm}\rangle$
can be diagonalized independently having the same eigenvalues and eigenvectors
of a 2 site Holstein model (see Appendix \ref{cf}). 
The hamiltonian matrix in the $S_{1/2}$ can be written

\beq
H_{1/2} = 	\left ( \begin{array}{cc}
			H_{1/2,+}+\omega_0\uni  \ax a&-g\uni (\ax+a)\\
			-g\uni(\ax+a)&H_{1/2,-}+\omega_0\uni  \ax a\\
			\end{array} 
		\right)
\label{H_ABC}
\eeq.

Here  $\uni$ is the 3$\times$ 3 unit matrix.
We can diagonalize $H_{1/2}$ in the phonon space by means of the 
unitary transformation 
\beq
U =	\frac{1}{2}\left ( \begin{array}{cc}
		(1+(-)^{\ax a})\uni&(-1+(-)^{\ax a})\uni\\
		(-1+(-)^{\ax a})\uni&(1+(-)^{\ax a})\uni\\
			\end{array} 
	\right).
\eeq
Using the property given in Eq. (\ref{UX}) we get for
$\tilde{H}_{1/2}=U H_{1/2} U^{-1}$
\beq
\tilde{H}_{1/2} =	\left ( \begin{array}{cc}
(\omega_0 \ax a -g(\ax+a))\uni+H_{1/2}^{at}-t(-)^{\ax a}\Delta&0\\
0&(\omega_0 \ax a +g(\ax+a))\uni+H_{1/2}^{at}+t(-)^{\ax a}\Delta\\
			\end{array} 
	\right)
\label{H_1_2}
\eeq
where we have splitted $H_{1/2,\pm}=H_{1/2}^{at}\pm t\Delta$ in the atomic 
($t=0$) part and in the hopping dependent term
\beq
\Delta=\left ( \begin{array}{ccc}
	0&1&0\\
	1&0&0\\
	0&0&1\\
	\end{array}
	\right )
\eeq
and $H_{1/2}^{at}$ can be obtained from Eq. (\ref{H_ABC}) with $t=0$.
The hamiltonian Eq. (\ref{H_1_2}) can be diagonalized in each spin sector by an
inddependent transformation which diagonalizes $H_{1/2,\pm}$ for a given phonon
number $n$. The diagonalization gives six eigenvalues $E_{1/2,\pm}(n)$ for each
phonon number. For $n=0$ and $J_H \gg t,J_1$ the lowest of them are
\beqa
E^{FE}_{1/2,-}&=&-\frac{J_H}{4}-t+\frac{J_1}{4}\\
E^{AF}_{1/2,-}&=&-\frac{J_H}{4}-\frac{t}{2}-\frac{J_1}{2}
\eeqa 
By replacing $t\rightarrow t(-)^{\ax a}$ we are left with a purely phononic
hamiltonian in the $FE/AF$ sector
\beqa
\tilde{H}^{FE}_{1/2}&=&\omega_0 \ax a -g(\ax+a)-\frac{J_H}{4}-t(-)^{\ax a}+\frac{J_1}{4}\\
\tilde{H}^{AF}_{1/2}&=&\omega_0 \ax a -g(\ax+a)-\frac{J_H}{4}-\frac{t}{2}(-)^{\ax a}-\frac{J_1}{2}
\eeqa
The comparison between the energies of the $FE$ and the $AF$ phases
leads to the condition (\ref{conds12}) for the magnetic transition.


\begin{references}
\bibitem{note-extrema} Notice that Eq. (\ref{theta_c}) gives the approximate 
$\theta_{MLF}$ of Ref. \cite{indiani} once the approximation 
$E_{kin}\simeq -t \exp{-alpha^2}$ valid in the extreme
anti-adiabatic regime has been made.
\end{references}
\end{document}